\documentclass[aps,preprint,prd]{revtex4}
\begin{document}
\draft
\title{Chiral SU(3) Quark Model Study of Tetraquark States: $cn\bar n\bar s/cs\bar s\bar s$}
\author{H. X. Zhang[1], W. L. Wang[2], Y.-B. Dai[3] and Z. Y. Zhang[1]}
\address{[1]Institute of High Energy Physics, P.O. Box 918-4, Beijing 100049, People's Republic of China\\
 [2]Institute of Particle Physics, Central China Normal University, Wuhan 430079, People's Republic of China\\
 [3]Institute of Theoretical Physics, P.O. Box 2735, Beijing 100080, People's Republic of China}
%\date{\today}

\begin{abstract}
The new members of the charm-strange family $D_{sJ}^{*}(2317)$,
$D_{sJ}(2460)$ and $D_s(2632)$, which have the surprising
properties, are challenging the present models. Many theoretical
interpretations have been devoted to this issue. Most of authors
suggest that they are not the conventional $c\bar s$ quark model
states, but possibly are four-quark states, molecule states or
mixtures of a P-wave $c\bar s$ and a four-quark state. In this work,
we follow the four-quark-state picture, and study the masses of
$cn\bar n\bar s/cs\bar s\bar s$ states ($n$ is $u$ or $d$ quark) in
the chiral SU(3) quark model. The numerical results show that the
mass of the mixed four-quark state ($cn\bar n\bar s/cs\bar s\bar s$)
with spin parity $J^P=0^{+}$ might not be $D_s(2632)$. At the same
time, we also conclude that $D_{sJ}^{*}(2317)$ and $D_{sJ}(2460)$
cannot be explained as the pure four-quark state.
\end{abstract}
\maketitle



Suggest heading: Hadronic Physics and QCD

\section{Introduction}

The discovery of several members of low-lying narrow charm-strange
meson in experiment has excited great interests of physicists. About
three years ago, BABAR Collaboration at SLAC reported the
observation of the narrow state $D_{sJ}^{*}(2317)$ [1], and it was
confirmed by CLEO Collaboration [2]. At the same time, another
narrow state $D_{sJ}(2460)$ was measured by this collaboration. Bell
Collaboration at KEK also confirmed these two resonances, and
studied their properties later [3], whose results are consistent
with the spin-parity assignments of $J^P=0^{+}$ for
$D_{sJ}^{*}(2317)$ and $J^P=1^{+}$ for $D_{sJ}(2460)$. Subsequently,
the SELEX Collaboration reported an evidence for a new puzzling
narrow state at a mass of $2632.6\pm 1.6MeV/c^2$ with total width
$\Gamma <17MeV$, which is known as $D_s(2632)$. It decays
preferentially to $D_s^{+}\eta $ rather than to $D^{+}K^0$ or
$D^0K^{+}$ [4]. All these new states have the same surprising
properties: (1) they are narrow; (2) their masses are smaller than
most theoretical predictions for $c\bar s$ states. Besides,
$D_{sJ}^{*}(2317)$ and $D_{sJ}(2460)$ have been observed in
isospin-violating decay channels. Their surprising properties have
attracted much attentions and various theoretical explanations were
proposed. Since their masses in the constituent quark model are much
higher than the experimental values, some people suggest that they
might constitute a new group which is different from the normal
members of the charm-strange family, namely they are not the simple
$c\overline{s}$ composition, but possibly are four-quark (4q)
states, molecule states or mixtures of 4q states and $c\overline{s}$
[5-10]. In this work, we follow the 4q-state picture.

According to the experiments, $D_{sJ}^{*}(2317)$ decays to
$D_s^{+}\pi ^0$ and its $J^P=0^{+}$, while $D_{sJ}(2460)$ decays to
$D_s^{*+}\pi ^0$ and $J^P=1^{+}$. Some people propose that $cn\bar
n\bar s$ component dominates in both two states, where $n$
generically represents either of $u$, $d$ quark [6]. As for
$D_s(2632)$, there are three kinds of suggestions for its quark
content. Some authors denote it as $cs\overline{s}\overline{s}$ [6,
7], others suggest the $cn\overline{n}\overline{s}$ content [8], and
Y.-R. Liu et al. [9] guess that $D_s(2632)$ is a 4q state in the
$SU(3)_F $ 15 representation with quark content $\frac
1{2\sqrt{2}}(ds \bar d+sd\bar d+su\bar u+us\bar u-2ss\bar s) \bar
c$. In these papers they only give the qualitative analysis for
these three narrow states, and in fact it is very indigent in
dynamical study. Thus, we think that it is necessary to study the 4q
states based on the QCD-inspired model.

Using the constituent SU(3) quark model, which is quite successful
in explaining the baryon spectrum, J. Vijande et al. give a
reasonable interpretation for $D_{sJ}^{*}(2317)$ and $D_{sJ}(2460)$
as the mixtures of a P-wave $c\bar s$ and a $cn\bar n\bar s$ state,
but they do not find any states around 2600 MeV [10]. The Chiral
SU(3) quark (CSQ) model [11] was proposed by generalizing the idea
of the SU(2) $\sigma$ mode to the flavor SU(3) case, in which the
nonet pseudo-scalar meson exchanges and the nonet scalar meson
exchanges are considered in describing the medium and long range
parts of the interaction, and the one-gluon-exchange (OGE) potential
is still retained to contribute the short range repulsion. Since the
CSQ model can reasonably reproduce not only the binding energy of
deuteron but also the nucleon-nucleon (NN) scattering phase shifts
of different partial waves and the hyperon-nucleon (YN) cross
sections by the resonating group method calculations [11, 12], we
try to extend this model to study the heavy-light 4q systems.

The CSQ model is based on the constituent quark model of the light
quark systems, in which the constituent mass appears because of the
vacuum spontaneous breaking, and at the same time the coupling
between Goldstone bosons and quarks is automatically introduced for
restoring the chiral symmetry and these boson exchanges are
essential to obtain a correct description of the $NN$ phenomenology
and the light baryon spectrum. But for the heavy quarks, their
current masses are very closed to the actual values, this means that
their constituent part is very small and the vacuum spontaneous
breaking effect is not important. Therefore, for the light quark
systems, the Goldstone boson exchanges are necessary to be
considered, but for the heavy quark systems, the one-gluon-exchange
interaction between quarks and the confinement potential are enough
to describe their main properties. This basic framework is
consistent with the QCD inspire and it is the usual treatment in
many works. For the heavy-light quark systems, whether the chiral
SU(3) quark model should be extended to chiral SU(4) by including D
mesons or not? In our opinion, at least as the first step, the
Goldstone boson exchange between light and heavy quarks is
unnecessary to be considered, because the coupling between heavy
quark and Goldstone boson is unimportant and the masses of D mesons
are quite large, located inside the chiral symmetry scale, which is
regarded about $(0.15-0.2fm)^{-1}$ usually. Therefore, Goldstone
boson exchanges will not be considered in the heavy-light quark
systems.

In this work, 4q states with quark content $cn\bar n\bar s$ and
$cs\bar s\bar s$ are studied. The parameters for the light quark
pairs are taken from our previous work, and for the heavy-light
quark pairs they are obtained from fitting the masses of $D$, $D^*$,
$D_s$, $D_s^*$, $\eta _c$, $J/\Psi$ and $h_1(1p)$, which will be
introduced detailedly in the theoretical framework. Considering
$D_s(2632)$ as a mixture of $cn\bar n\bar s$ and $cs\bar s\bar s$ 4q
states, we compute its mass which is about 90MeV higher than the
experimental value. Moreover, we show that $D_{sJ}^{*}(2317)$ and
$D_{sJ}(2460)$ cannot be explained as the 4q state in our model.

Finally, we repeat our calculation in the extended CSQ model [12]
which is proposed by extending the CSQ model to involve vector meson
exchanges. In addition, the OGE which dominantly governs the
short-range quark-quark (qq) or quark-antiquark ($q\bar q$)
interaction in the original CSQ model, is now nearly replaced by the
vector meson exchanges. Using this model, a challenging problem,
whether OGE or vector meson exchange is the right mechanism for
describing the short range part of the strong interaction, or both
of them are important, is studied. For the light quark systems, from
study of baryon-baryon scattering processes to the baryon-meson
systems in the extended CSQ model, some results are similar to those
given by the chiral unitary approach study [13]. Thus, in this
paper, we also try to calculate the heavy-light 4q states in this
extended model. We conclude that it is not suitable to study the
physics of heavy-light quark systems in the extended CSQ model.

The paper is arranged as follows. The theoretical framework of the
CSQ model, the determination of parameters and the wave functions
for $cn\overline{n}\overline{s}$ and $cs\overline{s}\overline{s}$
states are briefly introduced in Section II. The numerical results
are listed and discussed in Section III. Finally summary is shown in
Section IV.

\section{Theoretical framework}

\subsection{The model}

In our CSQ model, the Hamiltonian of a 4q system is written as
\begin{equation}
H=\sum_im_i+\sum_iT_i-T_G+\sum_{i<j=1}^4V_{ij}\; ,
%     (1)
\end{equation}
where $m_i$ is the mass of the $i$th quark, $T_G$ is the kinetic
energy operator for the c.m. motion, and $V_{ij}$ represents the
interactions between $qq$ or $q\bar q$. As for $qq$ pair,
$V_{ij}=V_{ij}^{conf}+V_{ij}^{OGE}+V_{ij}^{ch}$, where the
confinement potential $V_{ij}^{conf}$, which provides the
non-perturbative QCD effect in the long distance, is taken as linear
form in this work, namely
\begin{equation}
V_{ij}^{conf}=-(\lambda _i^c\cdot \lambda
_j^c)(a_{ij}r_{ij}+a_{ij}^0)\;  .
%   (2)
\end{equation}
Moreover, the expression of OEG potential $V_{ij}^{OGE}$ is
\begin{equation}
V_{ij}^{OGE}=\frac 14g_ig_j(\lambda _i^c\cdot \lambda _j^c)\left\{
\frac 1{r_{ij}}-\frac \pi 2(\frac 1{m_i^2}+\frac
1{m_j^2}+\frac{4(\sigma _i\cdot \sigma _j)}{3m_im_j})\delta (\vec
r_{ij})\right\} \; ,
%   (3)
\end{equation}
which governs the short-range perturbative QCD behavior.
$g_{i}g_{j}$ is the OGE coupling constant, which is an effective
form of the momentum-dependant quark-gluon coupling strength $\alpha
_s$, and it is frozen for each flavor sector. $V_{ij}^{ch}$
represents the interactions from chiral field coupling and describes
the non-perturbative QCD effect of the low-momentum medium-distance
range, which can be expressed as
\begin{equation}
V_{ij}^{ch}=\sum_{a=0}^8V_{s_a}(\vec
r_{ij})+\sum_{a=0}^8V_{ps_a}(\vec r_{ij})\;     ,
%    (4)
\end{equation}
where
\begin{equation}
V_{s_a}(\vec r_{ij})=-C(g_{ch},m_{s_a},\Lambda )(\lambda _i^a\cdot
\lambda _j^c)X_1(m_{s_a},\Lambda ,r_{ij})+V_{s_a}^{ls}(\vec
r_{ij})\; ,
%   (5)
\end{equation}
\begin{equation}
V_{ps_a}(\vec r_{ij})=C(g_{ch},m_{ps_a},\Lambda )(\lambda _i^a\cdot
\lambda _j^c)\frac{m_{ps_a}^2}{12m_{q_i}m_{q_j}}X_2(m_{ps_a},\Lambda
,r_{ij})(\sigma _i\cdot \sigma _j)+V_{ps_a}^{ten}(\vec r_{ij})\;,
%    (6)
\end{equation}
with
\begin{equation}
C(g_{ch},m,\Lambda )=\frac{g_{ch}^2}{4\pi }\frac{\Lambda ^2m}{\Lambda ^2-m^2}%
\;  ,
%     (7)
\end{equation}
\begin{equation}
X_1(m,\Lambda ,r)=Y(mr)-\frac \Lambda mY(\Lambda r)\;  ,
%      (8)
\end{equation}
\begin{equation}
X_2(m,\Lambda ,r)=Y(mr)-\left( \frac \Lambda m\right) ^3Y(\Lambda
r)\;   ,
%     (9)
\end{equation}
\begin{equation}
Y(x)=\frac 1xe^{-x}\;   ,
%   (10)
\end{equation}
and $m_{s_a}$ ($m_{ps_a}$) is the mass of the scalar (pseudoscalar)
meson.

The interactions for $q\bar q $ pair include two parts: direct
interaction and annihilation part
\begin{equation}
V_{q\bar q}=V_{q\bar q}^{dir}+V_{q\bar q}^{ann}\;  ,
%   (12)
\end{equation}
\begin{equation}
V_{q\bar q}^{dir}=V_{q\bar q}^{conf}+V_{q\bar q}^{OGE}+V_{q\bar
q}^{ch}\; .
%       (13)
\end{equation}
In this work, we neglect the contribution of annihilation part
firstly. The detailed expression of $V_{q\bar q}^{dir}$ can be
obtained from $V_{qq}$. As for $V_{q\bar q}^{conf}$ and $V_{q\bar
q}^{OGE}$, the transformation from $V_{qq}$ to $V_{q\bar q}$ is
given by $\lambda _i^c\cdot \lambda _j^c\rightarrow -\lambda
_i^c\cdot \lambda _j^{*c}$, while for $V_{q\bar q}^{ch}$ it is
$\lambda _i^a\cdot \lambda _j^a\rightarrow \lambda _i^a\cdot \lambda
_j^{*a}$.

Once perturbative (OGE) and non-perturbative (confinement and chiral
symmetry breaking) aspects of QCD have been considered, one ends up
with a $q_iq_j$ interaction of the form
\begin{equation}
V_{q_iq_j}=\left\{
\begin{array}{l}
q_iq_j=n\bar n,n\bar s,s\bar s,\bar n\bar s, \bar s\bar s
\Rightarrow V_{q_iq_j}^{conf}+V_{q_iq_j}^{OGE}+V_{q_iq_j}^{ch}
\\
q_iq_j=cn,cs, c\bar n, c\bar s \Rightarrow
V_{q_iq_j}^{conf}+V_{q_iq_j}^{OGE}\text{ }
\end{array}\right.
%   (11)
\end{equation}
Note that for the heavy-light quark pairs, the Goldstone boson
exchanges will not be considered.

\subsection{Determination of the parameters}

The interaction parameters include the OGE coupling constant $g_i$,
the confinement strengths $a_{ij}; a_{ij}^0$, and the chiral
coupling constant $g_{ch}$. In our 4q systems, there are three light
quarks (u, d, s) and one heavy quark (c). From Eq. (13), there are
three kinds of interactions in the light quark pairs, i.e. confining
potential, OGE potential and chiral potential, while for the
heavy-light quark pairs the latter one will not be considered.

(1). The parameters for the light quark pairs (Set I)

The parameters for the light quark pairs are taken from our previous
work [14], which gave a satisfactory description for the energies of
the baryon ground states, the binding energy of the deuteron
($B_{deu}$), the NN scattering phase shifts, and the YN cross
section. The procedure for parameters determination is listed in
Ref. [14, 15] in detail. For simplicity, we only show them in Table
I. The theoretical results for the energies of baryon ground states
and $B_{deu}$ are shown in Table II. In our calculation, $\eta$ and
$\eta ^{\prime }$ mesons are mixed by $\eta_1$ and $\eta_8$, the
mixing angle $\theta _{ps}$ is taken to be the usual value with
$\theta _{ps}=-23^{0}$. Here the scalar meson mixing is not
considered, i.e. $\theta _s=0$. The coupling constant for scalar and
pseudo-scalar chiral field coupling, $g_{ch}$, is determined
according to the relation $\frac{g_{ch}^2}{4\pi }=\frac
9{25}\frac{g_{NN\pi }}{4\pi }\frac{m_u^2}{M_N^2}$.

\begin{table}
\caption{Model parameters for the light quark pairs (Set I). The
masses of u (d) and s quark: $m_u=m_d=313MeV$, $m_s=470MeV$. The
meson masses and the cut-off: $m_{\sigma ^{\prime }}=m_\epsilon
=m_\kappa =980MeV$, $m_\pi =138MeV$, $m_K=495MeV$, $m_\eta =549MeV$,
$m_{\eta ^{\prime }}=957MeV$, and $\Lambda =1100MeV$.}
\begin{center}
\begin{tabular}{ll}
\hline Model Parameters (Set I)  &   \\ \hline
$b_u$ (fm) & 0.5 \\
$m_\sigma $ (MeV) & 595 \\
$\theta _{ps}$ & $-23^{0}$ \\
$\theta _s$ & 0 \\
$g_{NN\pi}$ & 13.67 \\ \hline
$g_u$; $g_s$ & 0.886; 0.917 \\
$a_{uu}$ (MeV/fm) & 90.41 \\
$a_{uu}^0$ (MeV) & -79.66 \\
$a_{us}$ (MeV/fm) & 104.2 \\
$a_{us}^0$ (MeV) & -76.19 \\
$a_{ss}$ (MeV/fm) & 155.3 \\
$a_{ss}^0$ (MeV) & -86.70 \\ \hline
\end{tabular}
\end{center}
\end{table}

\begin{table}
\caption{The energies of the baryon ground states and the binding
energy of deuteron.}
\begin{center}
\begin{tabular}{lccccccccc}
\hline & N & $\Sigma $ & $\Xi $ & $\Lambda $ & $\Delta $ & $\Sigma
^{*}$ & $\Xi ^{*} $ & $\Omega $ & $B_{deu}$ \\ \hline
Exp. (MeV) & 939 & 1194 & 1319 & 1116 & 1232 & 1385 & 1530 & 1672 & 2.224 \\
Theor. (MeV) & 939 & 1194 & 1334 & 1116 & 1237 & 1375 & 1515 & 1657 & 2.13 \\
\hline
\end{tabular}
\end{center}
\end{table}

(2). The parameters for the heavy-light quark pairs (Set II)

In order to obtain the OGE coupling constant $g_c$ in the
heavy-light quark pairs, we refer to the effective scale-dependent
strong coupling constant in Ref. [10], i.e.
\begin{equation}
\alpha _s(\mu _{ij})=\frac{\alpha _0}{\ln \left[ (\mu _{ij}^2+\mu
_0^2)/\Lambda _0^2\right] }\;   ,
%      (14)
\end{equation}
where $\mu _{ij}$ is the reduced mass of the $qq$ ($q\bar q$) system
and $\alpha _0$, $\mu _0$ and $\Lambda _0 $ are fitted parameters.
$\alpha _s(\mu _{ij})$ is fitted to the behavior of the standard
expression for the running coupling constant $\alpha _s(Q^2)$. In
Eq. (14) the typical momentum scale of each flavor sector is
assimilated to the reduced mass of the system. Because $g_ig_j$ is
also an effective form of $\alpha _s$, according to the numerical
results of $\alpha _s(\mu _{cc})$, $\alpha _s(\mu _{cu})$ and
$\alpha _s(\mu _{cs})$ we can estimate a rough range for $g_c$:
$0.53\sim 0.6$ with $m_c= 1300\sim 1700MeV$.

In the case of the confinement strengths $a_{cu};a_{cu}^{0}$,
$a_{cs};a_{cs}^{0}$ and $a_{cc};a_{cc}^{0}$, as a preliminary
sample, we fix them by fitting the masses of $D$, $D^{*}$, $D_s$,
$D_s^{*}$, $\eta _c$, $J/\Psi$ and $h_c(1p)$. In order to avoid an
unbound spectrum, the delta-function in OGE potential has to be
regularized, namely
\[
\delta (\vec r_{ij})\Rightarrow \frac 1{4\pi }\Lambda
_{ij}^3Y(\Lambda _{ij}r_{ij})\;     ,
\]
where $Y(x)=e^{-x}/x$, and $\Lambda _{ij}=\mu _{ij}/\hat{r}_0$ that
follows by the flavor-dependent regularization $r_0(\mu )$ in Ref.
[10]. In this work, we choose $g_c=0.58$ and $m_c$=1430MeV to
calculation. The numerical results of the confinement strengths and
masses of $D$, $D^{*}$, $D_s$, $D_s^{*}$, $\eta _c$, $J/\Psi$ and
$h_c(1p)$ are shown in Table III and IV, respectively. From Table
IV, it is seen that the masses of these states are reasonably
consistent with the experimental values.

\begin{table}
\caption{Model parameters for the heavy-light quark pairs (Set II).
$m_{c}=1430MeV$.}
\begin{center}
\begin{tabular}{ll}
\hline Model Parameters (Set II)  & \\ \hline
$a_{cu}$ (MeV/fm) & 275 \\
$a_{cu}^0$ (MeV) & -155.9 \\
$a_{cs}$ (MeV/fm) & 275 \\
$a_{cs}^0$ (MeV) & -124.7 \\
$a_{cc}$ (MeV/fm) & 275 \\
$a_{cc}^0$ (MeV) & -77.8 \\\hline
\end{tabular}
\end{center}
\end{table}

\begin{table}
\caption{The masses of $D$, $D^*$, $D_s$, $D_s^*$, $\eta _c$,
$J/\Psi$ and $h_c(1p)$. Experimental data are taken from PDG.}
\begin{center}
\begin{tabular}{lccccccc}
\hline Mesons & $D$ & $D^{*}$ & $D_s$ & $D_s^{*}$ & $\eta _c$ & $J/\Psi$ & $h_c(1p)$ \\
\hline
Exp. (MeV)  & 1867.7 & 2008.9 & 1968.5 & 2112.4 & 2979.6 & 3096.916 & 3526.21\\
Theor. (MeV) & 1888 & 2009 & 1969 & 2130 & 2990 & 3098 & 3568\\
\hline
\end{tabular}
\end{center}
\end{table}

Summing up the parameters for the light quark pairs and the
heavy-light quark pairs, the parameter group for 4q states
calculation is Set I+Set II.

\subsection{The wave functions of 4q states: ($cn\bar n\bar s/cs\bar s\bar s$)}

As for a 4q state ($q_1q_2\bar q_3\bar q_4$), the wave function will
be a tensor product of a spatial ($\phi ^S$), flavor ($\phi ^F$),
spin ($\chi _s$) and color ($\chi _C$) wave functions
\[
|\Psi _{q_1q_2\bar q_3\bar q_4}>=\phi ^S(q_1q_2\bar q_3\bar q_4)\phi
^F(q_1q_2\bar q_3\bar q_4)\chi _s(q_1q_2\bar q_3\bar q_4)\chi
_C(q_1q_2\bar q_3\bar q_4).
\]
With respect to the color wave function, one can couple the two
quarks and the two antiquarks to a color singlet 4q state in
different ways: (1) $\{|1_{13},1_{24}>, |8_{13},8_{24}>\}$, (2)
$\{|1_{14},1_{23}>, |8_{14},8_{23}>\}$ and (3) $\{|\bar
3_{12},3_{34}>, |6_{12},\bar 6_{34}>\}$. As for the 4q states
($cn\bar n\bar s$ and $cs\bar s\bar s$) we study in this paper, the
wave function of antiquark pair ($\bar n\bar s$ or $\bar s\bar s$)
should be antisymmetric in the CSQ model. The expression (3) is more
convenient to treat the anti-symmetrization of the two antiquarks.
Thus, we start from expression (3), then consider the configuration
mixing between the states with the same quantum numbers.

Let us describe the wave function of a 4q state as the following
form
\[
\Psi (4q)=\psi _{4q}(0s^4)\left[ (qq)_{I_1,S_1,C_1};(\bar q\bar
q)_{I_2,S_2,C_2}\right] _{I,S,(00)^C}\;    ,
\]
where $\psi _{4q}(0s^4)$ is the orbital part and all the four quarks
are in S-wave state, $\left[ (qq)_{I_1,S_1,C_1};(\bar q\bar
q)_{I_2,S_2,C_2}\right] _{I,S,C}$ is the flavor-spin-color part and
the wave function of antiquark pair $(\bar q\bar q)_{I_2,S_2,C_2}$
should be antisymmetric. For example, if the $\bar q\bar q$ pair is
symmetric in flavor ($(02)^F$), due to the Pauli principle the
color-spin wave function of the pair should be antisymmetric, i.e.
antisymmetric in color ($C_2=(10)^C$), symmetric in spin ($S_2=1$)
or symmetric in color ($C_2=(02)^C$), antisymmetric in spin
($S_2=0$). For the 4q wave function, the isospin-spin-color quantum
numbers ($I_1,S_1,C_1$) for the $qq$ pair and ($I_2,S_2,C_2$) for
the $\bar q\bar q$ pair should be combined to form a color singlet
($I,S,(00)^C$). Therefore, as for $cn\bar n\bar s$ with quantum
numbers $(J^P;I)=(0^{+};0)$, the possible configurations are read
\begin{equation}
\Psi _A=\psi _{cn\bar n\bar s}(0s^4)\left[ (cn)_{\frac
12,0,(01)^C};(\bar n\bar s)_{\frac 12,0,(10)^C}^{(10)^F}\right]
_{0,0,(00)^C}\;   ,
%        (16)
\end{equation}
\begin{equation}
\Psi _B=\psi _{cn\bar n\bar s}(0s^4)\left[ (cn)_{\frac
12,1,(20)^C};(\bar n\bar s)_{\frac 12,1,(02)^C}^{(10)^F}\right]
_{0,0,(00)^C}\;   ,
%       (17)
\end{equation}
\begin{equation}
\Psi _C=\psi _{cn\bar n\bar s}(0s^4)\left[ (cn)_{\frac
12,0,(20)^C};(\bar n\bar s)_{\frac 12,0,(02)^C}^{(02)^F}\right]
_{0,0,(00)^C}\;     ,
%           (18)
\end{equation}
\begin{equation}
\Psi _D=\psi _{cn\bar n\bar s}(0s^4)\left[ (cn)_{\frac
12,1,(01)^C};(\bar n\bar s)_{\frac 12,1,(10)^C}^{(02)^F}\right]
_{0,0,(00)^C}\;  ,
%           (19)
\end{equation}
where $(\bar n\bar s)^{(10)^F}=\frac 1{\sqrt{2}}(\bar n\bar s-\bar
s\bar n)$, and $(\bar n\bar s)^{(02)^F}=\frac 1{\sqrt{2}}(\bar n\bar
s+\bar s\bar n)$. Similarly, the wave functions for $cs\bar s\bar s$
states are
\begin{equation}
\Psi _E=\psi _{cs\bar s\bar s}(0s^4)\left[ (cs)_{0,0,(20)^C};(\bar
s\bar s)_{0,0,(02)^C}\right] _{0,0,(00)^C}\;    ,
%         (20)
\end{equation}
\begin{equation}
\Psi _F=\psi _{cs\bar s\bar s}(0s^4)\left[ (cs)_{0,1,(01)^C};(\bar
s\bar s)_{0,1,(10)^C}\right] _{0,0,(00)^C}\;    .
%        (21)
\end{equation}
As for other quantum numbers, such as $(J^P;I)=(0^{+};1)$,
$(1^{+};0)$ and $(1^{+};1)$, the wave functions can be written out
following the same rule. For saving space, we will not show them in
here.

Of cause, it is necessary to re-couple the color basis from $\{|\bar
3_{12},3_{34}>, |6_{12},\bar 6_{34}>\}$ to $\{|1_{13},1_{24}>,
|8_{13},8_{24}>\}$ to calculate the interactions of $q \bar q$ pair
in our calculation. At the same time, the re-coupling expression is
useful to make the physical meaning of the state obviously. For
instance,
\begin{eqnarray}
\Psi _A &=&-\frac 1{2\sqrt{6}}\psi _{c\bar nn\bar s}(0s^4)[(c\bar n)_{\frac 12,0,(00)^C};%
(n\bar s)_{\frac 12,0,(00)^C}]_{0,0,(00)^C}  \nonumber\\
&&-\frac 1{2\sqrt{6}}\psi _{c\bar sn\bar n}(0s^4)[(c\bar s%
)_{0,0,(00)^C};(n\bar n)_{0,0,(00)^C}]_{0,0,(00)^C}  \nonumber\\
&&-\frac 1{2\sqrt{2}}\psi _{c\bar nn\bar s}(0s^4)[(c\bar n%
)_{\frac 12,1,(00)^C};(n\bar s)_{\frac 12,1,(00)^C}]_{0,0,(00)^C}  \nonumber\\
&&-\frac 1{2\sqrt{2}}\psi _{c\bar sn\bar n}(0s^4)[(c\bar s%
)_{0,1,(00)^C};(n\bar n)_{0,1,(00)^C}]_{0,0,(00)^C}  \nonumber\\
&&+\frac 1{2\sqrt{3}}\psi _{c\bar nn\bar s}(0s^4)[(c\bar n%
)_{\frac 12,0,(11)^C};(n\bar s)_{\frac 12,0,(11)^C}]_{0,0,(00)^C}  \nonumber \\
&&+\frac 1{2\sqrt{3}}\psi _{c\bar sn\bar n}(0s^4)[(c\bar s%
)_{0,0,(11)^C};(n\bar n)_{0,0,(11)^C}]_{0,0,(00)^C}  \nonumber\\
&&+\frac 12\psi _{c\bar nn\bar s}(0s^4)[(c\bar n)_{\frac
12,1,(11)^C};(n\bar s)_{\frac 12,1,(11)^C}]_{0,0,(00)^C}  \nonumber\\
&&+\frac 12\psi _{c\bar sn\bar n}(0s^4)[(c\bar s%
)_{0,1,(11)^C};(n\bar n)_{0,1,(11)^C}]_{0,0,(00)^C},
\end{eqnarray}
where the first four rows in $\Psi _A$ describe physics mesons, and
the last ones are the color octet $q\bar q$ sector.

The trail wave function can be written as an expression of the 4q
state with several different harmonic oscillator frequencies $\omega
_i$,
\[
\psi _{4q}=\sum_i^n\alpha _i\phi _{4q}(b_i)\text{,}
\]
where $b_i^2=\frac 1{m\omega _i}$. Using the variation method the
energies of these states are obtained.

\section{Results and discussion}

According to the experiments of $D_{sJ}^{*}(2317)$ and
$D_{sJ}(2460)$, a natural interpretation is that they are P-wave
$c\bar s$ quark states, and their spin-parity are $J^P=0^{+}$ and
$1^{+}$, respectively [16]. Now, if we regard them as 4q states,
according to their decay modes their quark contents are $cn\bar
n\bar s$. Because of their isospin-violating decay channel, we focus
our calculation on such quantum numbers: $(J^P;I)=(0^{+},0)$,
$(0^{+},1)$, $(1^{+},0)$ and $(1^{+},1)$. The rms radii
($\left\langle r\right\rangle$ in fm) and masses ($M$ in MeV) of
$cn\bar n\bar s$ states are calculated and the numerical results are
listed in the first row of Table V. From the first two numerical
values, the masses of $cn\bar n\bar s$ state with $J^P=0^{+}$ are
much higher than the mass of $D_{sJ}^{*}(2317)$ whether its isospin
is 0 or 1. Therefore, we conclude that it is difficult to explain
$D_{sJ}^{*}(2317)$ as a pure 4q state in our calculation. From the
last numerical values, the same conclusion can be obtained for
$D_{sJ}(2460)$. This conclusion is consistent with the generally
acceptable opinion about these two surprising charm-strange mesons.
Since there is no compelling evidence that $D_{sJ}^{*}(2317)$ and
$D_{sJ}(2460)$ are nonconventional meson states, it is still
possible to interpret them as $c\bar s$ states or the mixtures of
$c\bar s$ and 4q states.

\begin{table}
\caption{The rms radii and masses ($\left\langle r\right\rangle ;M$)
(fm; MeV) for $cn\bar n\bar s/cs\bar s\bar s$ states with quantum
numbers ($J^P;I$)=($0^{+};0$), ($0^{+};1$), ($1^{+};0$) and
($1^{+};1$).}
\begin{center}
\begin{tabular}{lcccc}
\hline 4q states& ($0^{+};0$)& ($0^{+};1$) & ($1^{+};0$) & ($1^{+};1$)  \\
\hline
$cs\bar s\bar s$& (0.428; 2753)& (0.445; 2782) & (0.457; 2882) & (0.488; 2925)\\
$cs\bar s\bar s$& (0.393; 3124)& --- & (0.370; 3065) & --- \\
$cn\bar n\bar s/cs\bar s\bar s$& (0.417; 2729)& --- & (0.445; 2870) & ---\\
\hline
\end{tabular}
\end{center}
\end{table}

$D_{s}(2632)$ was regarded as one of excellent candidates for the
tetraquark, and many theoretical interpretations have been devoted
to this issue. Some people suggest that the $cs\bar s\bar s$
component dominates in $D_{s}(2632)$, and others denote its quark
content as $cn\bar n\bar s$. However, most of works are the
qualitative analysis, not the dynamical calculation based on a
QCD-inspired model. We study the mass of $D_{s}(2632)$ as a mixture
of $cn\bar n\bar s$ and $cs\bar s\bar s$ states in the CSQ model.
According to the decay modes of $D_{s}(2632)$ both particles in the
final states are pseudo-scalar mesons, thus it is probably an
isoscalar. Namely, in this calculation we focus on
$(J^P;I)=(0^{+};0)$. Firstly, the masses of $cn\bar n\bar s$ and
$cs\bar s\bar s$ states are calculated where the configuration
mixing between the states with same quantum numbers is considered,
and the numerical results are shown in the first column of Table V.
It is seen that the masses of both states are higher than 2632MeV,
especially the mass of $cs\bar s\bar s$. Thus $D_{s}(2632)$ cannot
be explained as $cs\bar s\bar s$ or $cn\bar n\bar s$ component alone
in our calculation.

Is it possible that $D_{s}(2632)$ is a mixture of $cn\bar n\bar s$
and $cs\bar s\bar s$? And in the CSQ model whether both states can
mix together or not? Firstly, let us discuss the possible decay
modes of the mixed state. In principle, the allowed decay modes
depend on the relationship between the tetraquark mass and the sum
of the masses of the possible decay products. In the case of $cn\bar
n\bar s$ with $(J^P; I)=(0^+; 0)$, let us take $\Psi _A$ as an
example, from the first four rows of Eq. (21), the possible decay
products are $D+K$, $D_s+\eta (n\bar n)$, $D^*+K^*$ and
$D_s^*+\omega (n\bar n)$. Since $M_{cn\bar n\bar s}$ is larger than
$M_D+M_K$ or $M_{D_s}+M_{\eta}$ and smaller than $M_{D^*}+M_{K^*}$
or $M_{D_s^*}+M_{\omega}$, the possible decay modes of $cn\bar n\bar
s$ are $D+K$ and $D_s+\eta$. Similarly, we can obtain the decay mode
of $cs\bar s\bar s$ is $D_s+\eta (s\bar s)$. Thus, the mixed state
can decay into $D+K$ and $D_s+\eta$ which is the same as
$D_{s}(2632)$'s decay modes. Secondly, it is allowable that $cn\bar
n\bar s$ and $cs\bar s\bar s$ can mix together through
meson-exchange and annihilation mechanics in our model. Therefore,
$D_{s}(2632)$ can be regarded as a mixture of $cn\bar n\bar s$ and
$cs\bar s\bar s$. In this work, the contributions of $K$ and $\kappa
$ meson exchanges between $cn\bar n\bar s$ and $cs\bar s\bar s$
states are taken into account, firstly, and the contributions of
annihilation mechanics are neglected. The numerical results are
listed in Table V. It is seen that the mass of the mixed state is
2729MeV, and its rms radius is 0.417fm which is a reasonable and
acceptable size for heavy-light 4q systems. If we change $g_c$
little higher, for example, $g_c=0.6$, the mass will be down about
several MeV.

Up to now, there are many assignments to describe $D_{s}(2632)$,
however, the masses and peculiar strong branching fractions reported
for this state appear inconsistent with any of these assignments. In
this work, we give a preliminary estimation to the strong branching
fraction firstly. According to our calculation, if $D_{s}(2632)$ is
the mixture of $cn\bar n\bar s$ and $cs\bar s\bar s$, its mass is
2729MeV with about 95.8\% of $cn\bar n\bar s$ content. Namely,
$cn\bar n\bar s$ is the dominant component in $D_{s}(2632)$. As an
example, from $\Psi _A$ in Eq. (21), we know that the rates for
$cn\bar n\bar s$ decay into $D(c\bar n)+K(n\bar s)$ and
$D_s^{+}(c\bar s)+\eta (n\bar n)$ is equal and $n\bar n$ component
is a very small part in $\eta$ meson. Thus, it seems that the
$cn\bar n\bar s/cs\bar s\bar s$ mixed state should not dominantly
decay into $D_s^{+}\eta $. In other words, the strong decay
branching fraction in our calculation might also disagree with the
SELEX data. Since $D_{s}(2632)$ is not confirmed further by other
laboratories, as for the mixed state at a mass of 2729MeV with
$J^P=0^{+}$ which is obtained in our calculation, is it another
tetraquark state? This should be answered by advanced experiments.
Since the rms radii of heavy-light 4q states we study now is about
0.4fm, at such short distance the annihilation mechanism should be
considered. Besides, $c\bar s$ states might play a role in the
$D_{s}(2632)$. All these aspects, which will be considered in the
next step, might improve our calculation.

In addition, we also calculate the mass of $cn\bar n\bar s/cs\bar
s\bar s $ state with $(J^P;I)=(1^{+};0)$, and the numerical results
are listed in the third column of Table V. It is seen that the mass
of this state is 2870MeV and its rms radius is 0.445fm. Namely, as
for the vector partner of $D_{s}(2632)$, its mass is about 2870MeV
in our model, which is about one hundred MeV higher than the
predicted result in Ref. [6].

In order to improve our feeling on studying 4q states in the CSQ
model, we try several groups of parameters to repeat our
calculation. The parameters are all roughly consistent with the
masses of $D$, $D^{*}$, $D_s$, $D_s^{*}$, $\eta _c$, $J/\Psi $ and
$h_c(1p)$. They will bring some differences on the results of the 4q
states we studied, but the influences are not big and the
qualitative results are almost the same.

Finally, we try to study the 4q states in the extended CSQ model
[13] where besides the nonet pseudo-scalar meson fields and the
nonet scalar meson fields, the coupling among vector meson fields
with quarks is also considered. The masses we obtained are roughly
$200\sim 400$ MeV higher than the results in the CSQ model. The
reasons are as follows: in the extended CSQ model the OGE potential
which dominantly governs the short-range $qq$ ($q\bar q$)
interaction in the original CSQ model, is now nearly replaced by the
vector meson exchanges. In the light quark systems it can also give
a reasonable interpretation for physical processes. However, for the
heavy-light quark systems, the Goldstone boson exchanges do not
contributed. If the contribution of OGE potential in the heavy-light
quark pairs is almost neglected, only confining potential cannot
give a reasonable description to the interactions. Thus, it is not
suitable to study heavy-light 4q systems in the extended CSQ model.
Until now, the extended CSQ model can do well in light quark
systems. It seem that OGE potential or vector meson exchanges can
independently govern the short-range $qq$ ($q\bar q$) interaction.
However, besides the confinement potential, the contribution of OGE
is unique to describe the $qq$ or $q\bar q$ interactions as the
chiral symmetry is exactly broken. Which is the real
short-range-mechanism of all the $qq$ ($q\bar q$) interaction? We
think that it is also a open question in our future research.

\section{Summary}

In this work, we study the masses of $cn\bar n\bar s$ and $cs\bar
s\bar s$ 4q states in the CSQ model, and try to give a reasonable
interpretation to $D_{sJ}^{*}(2317)$, $D_{sJ}(2460)$ and
$D_{s}(2632)$. The numerical results show that $D_{sJ}^{*}(2317)$
and $D_{sJ}(2460)$ cannot be explained as the pure 4q state, which
is consistent with the general acceptable opinion about these two
surprising charm-strange mesons. We suggest that the mixed $cn\bar
n\bar s/cs\bar s\bar s$ 4q state with $(J^P;I)=(0^{+};0)$, which
mass is 2729MeV and about 90MeV higher than the experimental mass of
$D_{s}(2632)$, might be a tetraquark state. In addition, we note
that it is not suitable for studying the heavy-light quark systems
in the extended CSQ model.

In our present calculation, only the contribution of meson-exchange
between $cn\bar n\bar s$ and $cs\bar s\bar s$ state is considered.
Because of the small size of the 4q states the contribution of
annihilation mechanics should be taken into account. Besides,
followed by Ref. [10], P-wave $c\bar s$ state in these puzzling
charm-strange mesons can play a role. The effects of both two
aspects will be study in next step to improve our calculation.

So far $D_{s}(2632)$ is not confirmed by other laboratories, so the
future experimental studies will be crucial for understanding it.
The most important measurement, such as determination of the $J^P$
quantum numbers, could test the theoretical calculation.

\section{Acknowledgement}

We are in debt to Professor Y. W. Yu and Dr. F. Huang for
stimulating discussions. This work was supported in part by the
National Natural Science Foundation of China: No. 10475087.

\end{document}